# Discharge structure hierarchy of highly electronegative plasma at low pressure and quasi- cold ion approximation

Rui-Ji Tang[1], Shu-Xia Zhao[1]*, Yu Tian[2]

[1]Key Laboratory of Material Modification by Laser, Ion, and Electron Beams (Ministry of Education), School of Physics, Dalian University of Technology, Dalian, 116024, China

[2]Experimental training center, Dalian University of Science and Technology, Dalian, 116052, China

Correspondence: zhaonie@dlut.edu.cn

## Abstract

In this paper, the discharge structure of an Ar/SF$_6$ inductively coupled plasma (ICP) at the low pressure is investigated by mean of a fluid simulation at the quasi- cold ion approximation with the room temperature magnitude. It is found the structure is hierarchal, including (1) the *macroscopic* characteristics, the stratification and the parabola profile in the stratified core, (2) the *microscopic* characteristics, the double layer at the interface of stratification and the related acoustic wave that is damped, and (3) the *mesoscopic* characteristics, the density bump within the parabola profile. In the mesoscopic scale, the density bump in the parabola is given by the self-coagulation behavior at the conditions of ionic recombination loss chemical source (with *negative* magnitude) and free diffusion. By means of a dimensional analysis and the analogy between the ionic recombination chemical loss and an *effective* physical drift force, the dynamics of self-coagulation paves the way for the anions to achieve the Boltzmann's relation of them. In the macroscopic scale, the transport dominance of low pressure and the intrinsic high electronegativity condition of Ar/SF$_6$ plasma give rise to the stratification that divides the Ar/SF$_6$ discharge structure into an electronegative plasma region (i.e., internal *core*) and an electropositive plasma region (i.e., external *halo*) and in the core the profiles of ionic densities are parabolas as the theory predicted. In the microscopic scale and at the interface of core and halo, the simulated double layer is found to be formed by the anions and cations sheaths and the acoustic wave occurs due to the ionic beams generated by the double layer potential drop. The propagation of ionic vibration is strongly damped because the temperatures of cations and anions equal. The simulated double layer can be transformed as the dipole and capacitor models because of the chamber configuration and discharging mechanism of ICP, and the double-value properties of plasma edge potential and cations flux at the double layer are investigated for interpreting the early relevant analytical works. The plasma edge means the core and halo interface at which the electrically neutral plasma and electrically non-neutral double layer (i.e., dual-sheath) meet.

## I. Introduction

The low-pressure and radio-frequency (RF) electronegative plasma sources are widely used in the Si-based material etching processes and the functional thin-film deposition [1, 2]. In general, the capacitive [3] and inductive radio frequency plasma sources are applied, and the inductively coupled plasma [4, 5] (abbreviated as the ICP) is characterized by its low-pressure discharge, high plasma density, independent control of ionic flux and energy, and simple device. At present, the discharge structure of the plasma source characterizes the profiles of plasma species densities, and the related theories illustrate the underlying physics and chemistry that give rise to the profiles. The physical process involves ambipolar diffusion, and the chemistry involves the inelastic collisions of electron impact and reactions of heavy species. Moreover, the plasma collective interaction, the Boltzmann's balance, and the Bohm's sheath criterion play important roles in forming the steady-state density profiles. Owing to the presence of anions in the discharge, the physical and chemical processes occurring in the electronegative plasmas are complex and their discharge structures are hierarchical, thus different with the electropositive plasmas. Studies of the discharge structure are important since they help people deep understand the mechanisms occurring in the discharging plasmas, and more importantly are helpful for exciting the novel applications of them.

In the last century, the discharge structure of electronegative plasma was widely investigated based on the theoretical analysis and simple numerical calculation of fluid model. In Refs. [6-9], the stratification of the electronegative discharge structure into the electronegative core consisting of cations and anions (with a small number of electrons) and the electropositive halo consisting of cations and electrons (with tiny anions) were investigated, at the approximation of cold ions that neglects the ion diffusion at the condition, $T_i = 0$. Herein, $T_i$ is the ions temperature. In Ref. [10], the effect of finite ion temperature was included, and it was found that the central anions and cations core is preserved but the previously sharp boundary between the core and halo is smoothed. Besides, in Ref. [11] the Boltzmann's relation was assumed to be applicable to both the anions and electrons, and the ion density profile in the electronegative core is parabolic, calculated by the theory when neglecting the bulk ionic recombination loss. The electron density is known to be constant at the assumption of the Boltzmann's balances of anions and electrons. Temporarily, the assumption that the anions are in the Boltzmann's balance and the related parabola of ions and constant profile of electrons were harshly criticized by Franklin *et al* in Refs. [12-14], who instead supported the fact that the profiles of ions and electrons are similar. Through purely mathematic analysis, Bogdonov in Ref. [15] verified the similarity of ions and electrons profiles when the time scale of attachments is shorter than that of the anions ambi-polar diffusion, and the anions Boltzmann's balance is thus supposed to meet at the reverse condition. In Ref. [16], the assumption of anions Boltzmann's relation was verified by Hershkowitz *et al* through the experiment. It is seen that the precursor of anions Boltzmann's relation is still not clear, although it plays important role in establishing the early-stage discharge structure theory of electronegative plasma.

The above analytic works were based on the electrically neutral plasma approximation. When the Poisson's equation was added, the double layer appears at the interface of core and halo, as illustrated in Refs. [17-22]. Specially, in Refs. [19, 20], it was found that the plasma edge potential is double valued when the anions are cold enough, in the low-pressure collisionless electronegative discharge with the plasma approximation. When the strict charge neutrality was relaxed, the double layer was observed at the plasma edge at the condition that the flux associated with the low potential solution is less than that of the high potential solution. Herein, the plasma edge is meant the interface of core and halo at which the electrically neutral bulk plasma and the electrically non-neutral double layer (dual-sheath) meet. The collisionless fluid model indicates the famous Tonks-Langmuir model that is described in Refs. [23, 24]. Besides for the fluid model, the kinetic model was utilized to investigate the electronegative discharge structure in Ref. [21], aimed at relaxing the assumption of anions Boltzmann's balance. It was found in this reference, the double-layer stratified discharge was observed when the effective anions temperature is low enough, in agreement with the fluid model prediction that is based on the anions Boltzmann's balance. Besides for the fluid and kinetic models that are both collisionless, the effects of ion-neutral collisions on the discharge structure and on the double layer and flux were investigated in Ref. [22], and it was found that the typical three types of discharge predicted by the collisionless models above, i.e., stratified structure without double layer, double layer stratified structure, and the uniform structure without stratification, were still predicted when the collisions are included. Moreover, the flux and potential at the collisional plasma edge hold the same characteristics as in the collisionless plasma. It is noticed the diffusion was not considered and the ions are drifted in the above analytic works of fluid model [19, 20, 22], and the influence of collisions on the discharge structure is hence ascribed to the influence of collision on the ambi-polar diffusion potential. Furthermore, the occurrence of double layer in the above fluid models is equivalent to the occurence of vibration in the attached halo. Nevertheless, when the ions diffusion at finite ions temperature was included the vibration disappears, as illustrated in Ref. [25], which was hence recognized as an artefact of fluid model that adopts the unphysical assumption of significant difference between cations and anions temperatures. It is seen that the double layer mechanism and related dynamics are not clear yet. In the present article, the double layer at the interface of core and halo is found to be accompanied by the damped acoustic wave.

The early analytic works of electronegative discharge structure [6-25] are less self-consistent, e.g., neglecting the energy deposition process and utilizing constant electron temperature, and the studies on the structure of electronegative plasma are thereby less complete. Moreover, in Ref. [26], though the energy deposition process of an ICP $SF_6$ discharge was considered in an elaborate fluid model, it is assumed to be spatially uniform. The simulated discharge profile was found to evolve from the parabola to the ellipse at increasing the pressure, as predicted by the theory of Ref. [27]. It is stressed that although the basic evolution of discharge structure from the parabola to the ellipse is given by the detailed fluid model, the discharge hierarchy was not displayed due to the uniform assumption of the ICP energy deposition process. Moreover, since the plasma approximation is assumed in the fluid model of Ref. [26], the stratification and double layer were not predicted either. Our fluid simulation of an Ar/$SF_6$ ICP source at the low pressure that couples the Poisson's equation and a self-consistent treatment of ICP power deposition is shown in this article. It is found the discharge structure is hierarchal and it includes the stratification, the parabola profile in the core, the double layer and related acoustic wave, and the self-coagulation. As we found already, the self-coagulation is the dominant behavior in the electronegative discharge [28, 29] and it provides new insights for re-recognizing the anions Boltzmann's balance. One more advantage of the fluid simulation of present article is that it is carried out in the two-dimensional space while the early analytic works were conducted in the one-dimensional space, which is probably the reason that the self-coagulation behavior was not predicted. It is believed the space ought to be closed for the electronegative plasma evolving the self-coagulation dynamics. It is stressed that even with the above advantages, the fluid model we used in this article is not completely self-consistent since the temperatures of all heavy species that include the ions and neutrals are assumed to be in the room temperature value, 300 K. Via the above analysis, it is seen that the ions diffusion is automatically included at the finite ion temperature assumed. Since the ionic Ohm's heating scheme is excluded, the work of present article is called as the fluid simulation at a quasi- cold ion approximation, which needs to be compared with the more self-consistent fluid simulation in future, such as the hybrid plasma equipment model that is abbreviated as the HPEM and described in Ref. [30]. The quasi-cold ions approximation is chosen in this article for it satisfies the necessary condition of above analytic works the anions are cold enough as compared with the electrons, which is helpful for us to examine the detailed discharge structure hierarchy.

The remainder of this article is organized as follows. In Sec. (2.1), firstly, the fluid model of the Ar/$SF_6$ ICP source at the quasi cold ion approximation is described. Then, the related discharge structure theories, parabola and self-coagulation, and the linear wave theory are presented in Secs. (2.2-2.4), respectively. In Sec. 3 the results and discussion are given and in Sec. 4 the conclusion and further remarks are presented.

## II. Methodology

### (2.1) Fluid model of Ar/$SF_6$ ICP source

In this study, the fluid model is used to simulate the Ar/SF$_6$ mixed ICP source at the low pressure, 10 mTorr, at the fixed Ar versus SF$_6$ ratio, 9:1, and the fixed power, 300W. The simulated density profiles are explained based on the theories described in the next sections of methodology. Herein, a brief description of the fluid model is given and more details of it can be found in Ref. [29].

The equations of electron density and energy of fluid model are given as follows.

$$\frac{\partial n_e}{\partial t} + \nabla \cdot \mathbf{\Gamma}_e = R_e, \tag{1}$$

$$\frac{\partial n_\varepsilon}{\partial t} + \nabla \cdot \mathbf{\Gamma}_\varepsilon + \mathbf{E} \cdot \mathbf{\Gamma}_e = R_\varepsilon + P_{ohm}. \tag{2}$$

Herein, $n_e$ and $n_\varepsilon$ are the number density and energy density of electrons, respectively. $\mathbf{\Gamma}_e$ and $\mathbf{\Gamma}_\varepsilon$ are the corresponding fluxes of number density and energy density of electrons, respectively. $R_e$ and $R_\varepsilon$ are the corresponding source terms for the number density and energy density of electrons, respectively. $P_{ohm}$ is the deposited power density via the electron Ohm's heating scheme of RF field in the azimuthal direction, i.e., $P_{ohm} = \frac{1}{2}\text{Re}\left(\sigma_e |E_\theta|^2\right)$, in which $\sigma_e$ is the electrical conductivity of electrons and $E_\theta$ is the azimuthal RF field (see next). It is seen that the energy deposition process of ICP source in our selected fluid model is self-consistent. And, $\mathbf{E}$ is the electrostatic field vector in the radial and axial directions. For solving the electron equations, the reduced mobility of electrons, $\mu_e N_n = 8.23 \times 10^{23}$ (V$^{-1}$·m$^{-1}$·s$^{-1}$), is used and the other transport coefficients of electrons that are reported in Ref. [29], such as the electron diffusion coefficient and the electron energy diffusion and mobility coefficients, are calculated based on the mobility. It is noticed that the expression of electron energy equation of present fluid model version is different and the heat conduction coefficient of electrons is expressed by means of the energy diffusion and mobility coefficients of electrons.

The mass fraction equation is used to describe the mass transport of heavy species in Eq. (3).

$$\rho \frac{\partial w_k}{\partial t} = \nabla \cdot \mathbf{j}_k + R_k. \tag{3}$$

Herein, $\rho$ is the total mass density of heavy species and $w_k$ is the mass fraction of species $k$. $\mathbf{j}_k$ is the diffusive and drift flux of species $k$. It is noticed that the heavy species temperatures are assumed to be the room temperature, i.e., the quasi- cold ion approximation as mentioned in Sec. I. The diffusion coefficients of heavy species used in Eq. (3) are calculated through the Chapman-Enskog method [31] based on the Lennard-Johns potential [32] with the potential characteristic length and potential energy minimum parameters, $\sigma$ and $\varepsilon$. The mobilities of heavy species are calculated from the diffusion coefficients through the Einstein's relation.

To describe the azimuthal RF field in the reactor, the Maxwell's equations are combined to express the Ampere's law in Eq. (4).

$$(j\omega\sigma_e - \omega^2\varepsilon_0\varepsilon_r)\mathbf{A} + \nabla \times (\mu_0^{-1}\mu_r^{-1}\nabla \times \mathbf{A}) = \mathbf{J}_a. \quad (4)$$

Herein, $j$ is the imaginary unit and $\omega$ is the angular frequency of power source, expressed as $2\pi f$ at $f = 13.56 \text{MHz}$. $\varepsilon_0$ and $\varepsilon_r$ are the vacuum permittivity and the relative permittivity of dielectric window material (quartz), respectively. $\mu_0$ and $\mu_r$ are the vacuum permeability and the relative permeability of coil that is made of copper, respectively. $\mathbf{A}$ is the magnetic vector potential, from which the RF magnetic and electric fields are calculated as, $\mathbf{B} = \nabla \times \mathbf{A}, \mathbf{E}_{rf} = -\frac{\partial \mathbf{A}}{\partial t}$. When considering the azimuthal symmetry, only the azimuthal component of RF electric field, $E_\theta$, and the axial and radial components of RF magnetic field, $B_r, B_z$, need to be addressed. $\mathbf{J}_a$ is the applied coil current density and its magnitude is varied until the required power, 300 W, is achieved. $\sigma_e$ is the electron conductivity and expressed as, $\sigma_e = \frac{n_e q^2}{m_e(\nu_e + j\omega)}$. Herein, $\nu_e$ is the elastic collisions of electrons with neutrals.

The Poisson's equation is used to calculate the electrostatic field in Eq. (5).

$$\begin{aligned}\mathbf{E} &= -\nabla V, \\ \nabla \cdot \mathbf{D} &= \rho_V.\end{aligned} \quad (5)$$

Herein, $\mathbf{D}$ is the electric displacement vector, from which the electrostatic field can be deduced through the permittivity, and $\rho_V$ is the spatial charge density. Most of the boundary conditions used for the above equation is given in Ref. [29], and herein only the heavy species wall boundary conditions are described, i.e., $\Gamma_- = 0$, $\Gamma_+ = \left(\frac{\gamma_f}{1-\gamma_f/2}\right)\frac{1}{4}n_+\sqrt{\frac{8RT}{\pi M_W}} + \mu_+ n_+ E$, $\Gamma_n = \Gamma_+$, $\Gamma_{F_2} = \frac{1}{2}\Gamma_F$. Herein, $\Gamma_-$, $\Gamma_+$, and $\Gamma_n$ are the fluxes of anions, cations, and neutrals at the walls, respectively, and $\gamma_f$ is the sticking coefficient on the walls.

The Ar/SF$_6$ gas-phase chemistry includes the elastic collision, the excitation and deexcitation, the ionization, the direct and dissociative attachments, and the dissociation of the electron impact reactions. The heavy species reactions include the neutral and ionic recombination, the detachment, the Penning's ionization and the charge exchange. The surface kinetics of species include the recombination and the de-excitation. The ICP reactor consists of the vacuum chamber, the dielectric window and the plasma chamber. The radius of dielectric window and vacuum chamber is 14 cm, and the heights of them are 1 cm and 3 cm, respectively. The radius of plasma chamber is 15 cm and the height of it is 13 cm. To simulate the plasma processing technique, the substrate with a radius of 13 cm and a height of 4 cm is seated at the bottom center of plasma chamber. The two-turn coil is installed above the dielectric window and the radial locations of two turns are 8 cm and 10 cm, respectively. More detail about the fluid model of Ar/SF$_6$ ICP source, such as the gas phase reaction list, the surface kinetics, and the ICP configuration, can be found in Ref. [29]. As stated in Ref. [29], the electron-impact cross sections from the *lxcat* website database [33] based on the Maxwellian's electron energy distribution function and the direct Arrhenius' chemical reaction rates are used together, for constructing the Ar/SF$_6$ chemistry. Since there are many types of cations and anions in the Ar/SF$_6$ chemistry, such as SF$_6^-$, SF$_5^-$, SF$_4^-$, SF$_3^-$, SF$_2^-$, F$^-$, F$_2^-$, and SF$_5^+$, SF$_4^+$, SF$_3^+$, SF$_2^+$, SF$^+$, F$^+$, F$_2^+$, S$^+$, Ar$^+$, the summed cations and anions densities are presented. For the net source of anions, all attachments that give any species of the above anions, all the detachments that delete any species of the above anions, and all recombinations of cations and anions that delete any species of the above anions are added. It is noticed the negative source of anions in the self-coagulation dynamics is dominated by the recombination of cations and anions, rather than the detachments of anions through neutrals, different with the electronegative and capacitively coupled plasma, abbreviated as the CCP, that operates at relatively high pressures [34] in a unit of 100 mTorr.

**(2.2) Theory of core parabola [11, 27]**

The balances of flux and charge density in the electronegative plasma of core stratified are given in Eqs. (6, 7). Here, $\Gamma_+$, $\Gamma_-$, and $\Gamma_e$ are the fluxes of cations, anions, and electrons, respectively. $n_+$, $n_-$, and $n_e$ are the densities of the cations, anions, and electrons, respectively. The electronegativity $\alpha$ is defined in Eq. (8), i.e., the ratio between the anions density and electrons density. The fluxes of cations, anions, and electrons at the drift and diffusion approximations are given in Eqs. (9-11). Herein, $D_+$, $D_-$, and $D_e$ are the diffusion coefficients of the cations, anions, and electrons, respectively. $\mu_+$, $\mu_-$, and $\mu_e$ are the mobilities of the cations, anions, and electrons, respectively. At correlating Eqs. (6-11), the flux of cations is rewritten as a function of the quantities, $\mu_+, \mu_-, \mu_e$, $D_+, D_-, D_e$, $n_+, n_-, n_e$ and $\alpha$, in Eq. (12). Like the electropositive plasma, the ambipolar diffusion coefficient of electronegative plasma, $D_{a+}$, is introduced in Eq. (13). The cations flux is then re-expressed as the product of the newly introduced coefficient and the cations density gradient, as illustrated in Eq. (14).

$$\Gamma_+ = \Gamma_- + \Gamma_e, \tag{6}$$

$$n_+ = n_- + n_e, \tag{7}$$

$$\alpha = n_- / n_e, \tag{8}$$

$$\mathbf{\Gamma}_+ = -D_+ \nabla n_+ + n_+ \mu_+ \mathbf{E}, \tag{9}$$

$$\mathbf{\Gamma}_- = -D_- \nabla n_- - n_- \mu_- \mathbf{E}, \tag{10}$$

$$\mathbf{\Gamma}_e = -D_e \nabla n_e - n_e \mu_e \mathbf{E}, \tag{11}$$

$$\mathbf{\Gamma}_+ = -\frac{(\mu_e + \mu_- \alpha) D_+ + \mu_+ (1+\alpha) D_e (\nabla n_e / \nabla n_+) + \mu_+ (1+\alpha) D_- (\nabla n_- / \nabla n_+)}{\mu_e + \mu_- \alpha + \mu_+ (1+\alpha)} \nabla n_+, \tag{12}$$

$$D_{a+} = \frac{(\mu_e + \mu_- \alpha) D_+ + \mu_+ (1+\alpha) D_e (\nabla n_e / \nabla n_+) + \mu_+ (1+\alpha) D_- (\nabla n_- / \nabla n_+)}{\mu_e + \mu_- \alpha + \mu_+ (1+\alpha)}, \tag{13}$$

$$\mathbf{\Gamma}_+ = -D_{a+} \nabla n_+. \tag{14}$$

As seen above, this version of ambipolar diffusion coefficient in Eq. (13) is complicated. Therefore, reasonable approximations are needed to reduce its complexity and then clear physics can emerge. To achieve this goal, the huge difference between the electron and anion temperatures that characterizes the non-thermal equilibrium plasma is utilized. In Eq. (15), the parameter $\gamma$, is introduced, which is defined as the ratio between the electron and anion temperatures. Here, $T_e$ and $T_i$ are the electron and anion temperatures, respectively. It is noticed that the anion and cation are in thermal equilibrium; hence, their temperatures are the same. The Boltzmann's balance is then adopted for both the electrons and anions, and the relation between the electron and anion densities is found, as shown in Eq. (16), which is expressed by means of the relative changes in both the anion and electron densities, and the $\gamma$ parameter. The condition of electrical neutrality is then performed among the density gradients, as shown in Eq. (17). Utilizing the Eqs. (16) and (17), the ratios between the gradients of the electrons density and the cations density and between the gradients of the anions density and the cations density are obtained in Eq. (18). It is noted that the ratios of these density gradients are both expressed as a function of $\gamma$ and $\alpha$. In addition, the relations between the diffusion coefficients and mobilities of different species are given in Eq. (19), based on the Einstein's relation. Using the Eqs. (18) and (19), the complex ambipolar diffusion coefficient shown in Eq. (13) is firstly simplified in Eq. (20).

$$\gamma = T_e / T_i, \tag{15}$$

$$\frac{\nabla n_-}{n_-} = \gamma \frac{\nabla n_e}{n_e}, \tag{16}$$

$$\nabla n_+ = \nabla n_- + \nabla n_e, \tag{17}$$

$$\frac{\nabla n_e}{\nabla n_+} = \frac{1}{1+\gamma\alpha}, \frac{\nabla n_-}{\nabla n_+} = \frac{\gamma\alpha}{1+\gamma\alpha}, \tag{18}$$

$$\frac{D_-}{D_+} = \frac{\mu_-}{\mu_+}, \frac{D_e}{D_+} = \gamma\frac{\mu_e}{\mu_+}, \tag{19}$$

$$D_{a+} = D_+ \frac{(1+\gamma+2\gamma\alpha)\left(1+\alpha\frac{\mu_-}{\mu_e}\right)}{(1+\gamma\alpha)\left(1+\frac{\mu_+}{\mu_e}(1+\alpha)+\frac{\mu_-}{\mu_e}\right)}. \tag{20}$$

At present, the simplified ambipolar diffusion coefficient is a function of the quantities $D_+$, $\alpha$, $\gamma$, and the mobilities $\mu_+, \mu_-, \mu_e$. Then, the approximations between the mobilities of heavy cations and anions and the electron mobility, as illustrated in Eq. (21), are used to further simplify the coefficient in Eq. (22). The present article is focused on the highly electronegative plasma, and so if the electronegativity $\alpha$ is high enough, e.g., ~100, the coefficient is again simplified to be a constant, $2D_+$, as shown in Eq. (23).

$$\mu_-/\mu_e \ll 1, \mu_+/\mu_e \ll 1, \tag{21}$$

$$D_{a+} \cong D_+ \frac{1+\gamma+2\gamma\alpha}{1+\gamma\alpha}, \tag{22}$$

$$\alpha \gg 1, D_{a+} \cong 2D_+. \tag{23}$$

Still at the Boltzmann's balances of the electrons and anions, the relation between the anion and electron densities is given in Eq. (24), which is different with the Eq. (16) that is focused on the density gradient. Here, $n_{e0}$ and $n_{-0}$ are the fixed central densities of electrons and anions, respectively. In non-thermal equilibrium plasma, the electron temperature $T_e$ is high, with an amplitude of several electron volts; however, the anion temperature $T_i$ is low, which is approximately hundreds of Kelvin. Therefore, the value of $\gamma$ is of the order of 100. At such a high value of $\gamma$, the term on the right-hand side of Eq. (24), $\left(\frac{n_-}{n_{-0}}\right)^{1/\gamma}$, tends to one and the electrons density on the left side of it becomes spatially unchangeable, i.e., equal to $n_{e0}$. This process is expressed in Eq. (25).

$$\frac{n_e}{n_{e0}} = \left(\frac{n_-}{n_{-0}}\right)^{1/\gamma}, \tag{24}$$

$$\gamma \cong 100.0, \left(\frac{n_-}{n_{-0}}\right)^{1/\gamma} \cong 1.0, n_e \cong n_{e0}. \tag{25}$$

In Eq. (26), the cation continuity equation of the electronegative plasma before simplified is given. On the left side of the equation, the flux is expressed as the product of the ambipolar diffusion coefficient and the density gradient of the cation. On the right-hand side of the equation, the chemical sources include the ionization reactions that create the cations and the recombination reactions that deplete the cations. Here, $n_0$ is the density of the target neutral atom for the electron collisions. $K_{iz}$ and $K_{rec}$ are the rate coefficients for the ionization and recombination reactions, respectively. The simplified continuity equation is shown in Eq. (27) and the simplifications used are given in the next three aspects. Firstly, on the left side of the continuity equation, the simplified ambipolar diffusion coefficient of Eq. (23), $2D_+$, is used in the flux expression. Secondly, on the right side of the continuity equation, the simplified constant electron density of Eq. (25), $n_{e0}$, is used in the ionization chemical term. Thirdly, the recombination loss of cations is neglected in the chemical source terms, which is validated at the low pressure, as mentioned in Sec. I. The simplified cation continuity equation of Eq. (27) can be analytically solved and the solution of it is a parabolic function, as shown in Eq. (28). Herein, the $\alpha_0$ parameter defines the electronegativity of the center, as shown in Eq. (29), and $l$ is the nominal position where the electronegativity is zero. Utilizing the Eqs. (7, 25), the cations density function of Eq. (28) can be expressed into the electronegativity function, as shown in Eq. (30).

$$-\frac{d}{dx}\left(D_{a+}\frac{dn_+}{dx}\right) = K_{iz}n_0 n_e - K_{rec}n_+ n_-, \tag{26}$$

$$-2D_+ \frac{d^2 n_+}{dx^2} = K_{iz}n_0 n_{e0}, \tag{27}$$

$$\frac{n_+}{n_{e0}} = \alpha_0\left(1-\frac{x^2}{l^2}\right)+1, \tag{28}$$

$$\alpha_0 = n_{-0}/n_{e0}, \tag{29}$$

$$\alpha = \alpha_0\left(1-\frac{x^2}{l^2}\right). \tag{30}$$

## (2.3) Theory of self-coagulation [28, 29]

The self-coagulation related theories of anions have been described in Refs. [28, 29] and herein a brief introduction of them is given. The steady-state continuity equation of anions that consists of the free diffusion flux and the negative source term given by the recombinations is expressed in Eq. (31).

$$-D_-\nabla^2 n_- = -n_- n_+ k_{rec} = -n_- v_{rec}. \tag{31}$$

Introducing the parameter, $k_- = \sqrt{\dfrac{v_{rec.}}{D_-}}$, into Eq. (31), the quasi-Helmholtz equation is obtained, as shown in Eq. (32).

$$\nabla^2 n_- - n_- \frac{v_{rec.}}{D_-} = \nabla^2 n_- - n_- k_-^2 = \nabla^2 n - n k^2 = 0. \tag{32}$$

For simplicity, the quantity subscripts are all removed in Eq. (32). In Eq. (33), this quasi-Helmholtz equation is reformed by the method of separation of variables in the cylindrical coordinate system at the azimuthal symmetry assumption.

$$\begin{aligned}
&\frac{1}{\rho}\frac{\partial}{\partial \rho}\left(\rho \frac{\partial n}{\partial \rho}\right) + \frac{\partial^2 n}{\partial z^2} - k^2 n = 0, \\
&n(\rho, z) = R(\rho)Z(z), \\
&Z'' + v^2 Z = 0, \\
&\frac{d^2 R}{d\rho^2} + \frac{1}{\rho}\frac{dR}{d\rho} - \left(k^2 + v^2\right)R = 0.
\end{aligned} \tag{33}$$

Herein, $v^2$ represents the eigenvalues. Utilizing the homogeneous boundary conditions of the above axial equation, the eigenvalues, $v^2$, and the related eigenfunctions, $Z_m(z)$, are acquired in Eq. (34).

$$\begin{aligned}
&v_m^2 = m^2 \pi^2 / l^2, \\
&Z_m = \sin(m\pi z / l), \\
&Z = \sum_{m=0}^{\infty} c_m Z_m = \sum_{m=0}^{\infty} c_m \sin(m\pi z / l).
\end{aligned} \tag{34}$$

As seen, the above radial equation is zero-order *imaginary* Bessel's equation because of the negative source. Since the density value is limited at the center, the imaginary Bessel's function, rather than the Hankel's function, is selected. The expression for $R(r)$ is then obtained in Eq. (35).

$$R = d_m I_0\left(\sqrt{k^2 + v_m^2}\,\rho\right) = d_m I_0\left(\sqrt{k^2 + m^2\pi^2/l^2}\,\rho\right). \tag{35}$$

The expression for $n(\rho,z)$, which is a product of $R(\rho)$ and $Z_m(z)$ is given in Eq. (36).

$$n(\rho,z) = R(\rho)Z(z) = \sum_{m=0}^{\infty} c_m \sin(m\pi z/l) \cdot d_m I_0(\sqrt{k^2 + v_m^2}\,\rho)$$
$$= \sum_{m=0}^{\infty} a_m \sin(m\pi z/l) \cdot I_0(\sqrt{k^2 + m^2\pi^2/l^2}\,\rho). \tag{36}$$

It is seen from Eq. (37) the delta distribution unrelated to the spatial coordinates evolves at using the mathematic limit ideas to the obtained density, which indicates the physics of self-coagulation, actually. More details about the self-coagulation theory can be found in Refs. [28, 29].

$$\begin{aligned}
n(\rho,z) &= R(\rho)Z(z) = \sum_{m=0}^{\infty} a_m \sin(m\pi z/l) \cdot I_0(\sqrt{k^2 + m^2\pi^2/l^2}\,\rho) \\
&= \lim_{m \to \infty}\left[a_m \sin(m\pi z/l) \cdot \infty\right] = \lim_{m \to \infty}\left[a_m \sin(m\pi z/l) \cdot \lim_{z \to 0}\frac{1}{z}\right] \\
&= \lim_{z \to 0}\left[\lim_{m \to \infty} a_m \sin(m\pi z/l) \cdot \frac{1}{z}\right] = \lim_{z \to 0}\left[\lim_{m \to \infty} a_m \cdot \frac{\sin(m\pi z/l)}{z\pi/l} \cdot \frac{\pi}{l}\right] \\
&= \lim_{\zeta \to 0}\left[\lim_{m \to \infty} a_m' \cdot \frac{1}{\pi}\frac{\sin(m\zeta)}{\zeta}\right] = \lim_{\zeta \to 0}\left[a_\infty' \lim_{m \to \infty}\frac{1}{\pi}\frac{\sin(m\zeta)}{\zeta}\right] \\
&= a_\infty' \lim_{\zeta \to 0}\delta(\zeta).
\end{aligned} \tag{37}$$

## (2.4) Dispersion relation of acoustic wave in the ionic pair plasma [35]

The momentum equation of cations in the ionic pair plasma or the strongly electronegative plasma is expressed as,

$$Mn\left[\frac{\partial \mathbf{v}_i}{\partial t} + (\mathbf{v}_i \cdot \nabla)\mathbf{v}_i\right] = en\mathbf{E} - \nabla p = -en\nabla\phi - \gamma_i KT_i \nabla n. \tag{38}$$

Herein, $M$ is the mass of cations, $n$ is the density of cations, and $\mathbf{v}_i$ is the velocity of cations, respectively. $\mathbf{E}$ and $\phi$ are related to the electrostatic field and potential. $p$ and $\gamma_i$ are the pressure of cations and the related adiabatic coefficient. $K$ is the Boltzmann's constant. The linear wave theory at the first order approximation gives the relation below related to the momentum equation.

$$-i\omega M n_0 v_{i1} = -en_0 ik\phi_1 - \gamma_i KT_i ik n_1. \tag{39}$$

Herein, $v_{i1}$, $\phi_1$, and $n_1$ are the vibration amplitudes of acoustic wave for the velocity, potential and density. $n_0$ is the background and uniform density of cations. $\omega$ and $k$ are the angular frequency and the wave number of acoustic wave, respectively, and $i$ is the imaginary unit. The continuity equation of cations in the ionic pair plasma or the strongly electronegative plasma is expressed as,

$$\frac{\partial n}{\partial t} + \nabla \cdot (n\mathbf{v}_i) = 0. \quad (40)$$

The linear wave theory at the first order approximation gives the relation below related to the continuity equation.

$$i\omega n_1 = n_0 i k v_{i1}. \quad (41)$$

At the conditions of the plasma electric neutrality and the Boltzmann's balance of anions, the following relation is obtained.

$$n_1 = n_0 \frac{e\phi_1}{KT_-} = n_0 \frac{e\phi_1}{KT_i}. \quad (42)$$

Herein, the anions temperature equals to the ion temperature at the quasi- cold ion approximation. Correlating the Eqs. (39, 41, 42), the dispersion relation of acoustic wave in the ionic pair plasma is obtained below, which equals to the speed velocity. The dispersion relation of linear wave theory and the Landau's damp will be used to explain the acoustic vibration in Sec. (3.3).

$$\frac{\omega}{k} = \left(\frac{KT_- + \gamma_i KT_i}{M}\right) \equiv v_s. \quad (43)$$

# III. Results and analysis
## (3.1) Stratification, parabola core, and ambi-polar self-coagulation

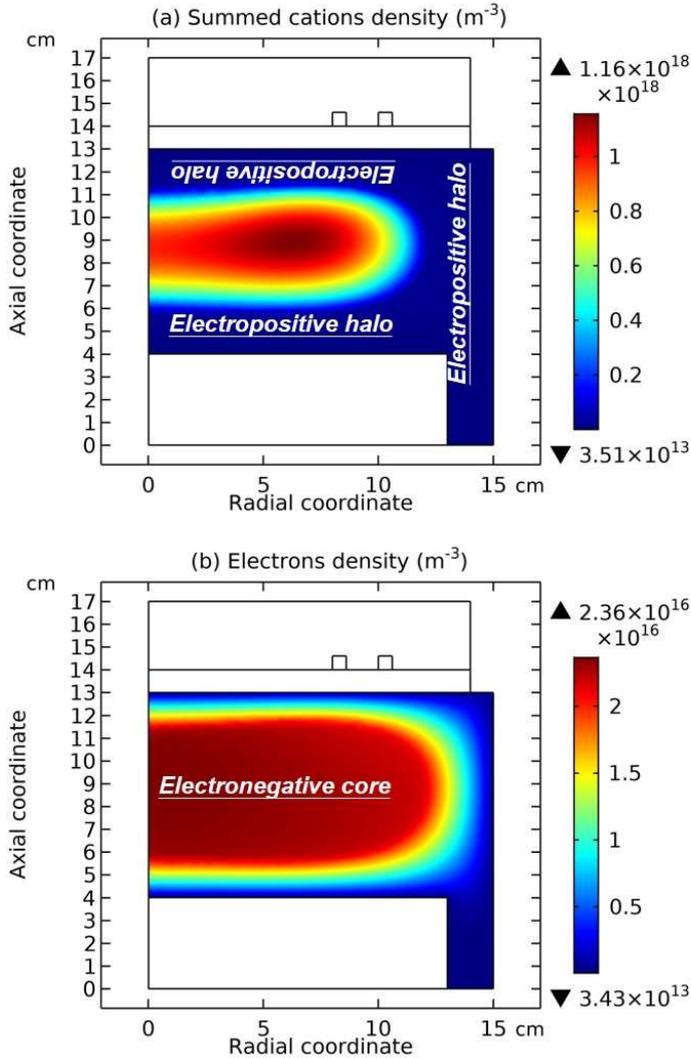

Figure 1. Simulated two-dimensional profiles of the summed cations density (a) and the electrons density (b), by the fluid model of Sec. (2.1). The discharge conditions of Ar/SF$_6$ ICP are 300W, 10mTorr and 10% SF$_6$ content and the simulated time is $0.1\,\text{s}$.

The simulated two-dimensional profiles of the summed cations density and the electrons density by the fluid model of Sec. (2.1) are presented in Fig. 1, and the axial and radial density profiles of the cations, the anions (also summed), and the electrons are given in Fig. 2. The discharge conditions of the Ar/SF$_6$ ICP source are 300 W, 10mTorr and 10% SF$_6$ content, and the simulated time is 0.1 s. It can be seen in Fig. 1 that the entire discharge structure is stratified into an electronegative core and an electropositive halo, and in the electronegative core the density axial profiles of the summed cations and anions in Fig. 2(a) are both parabolic, as predicted by the analytical theory of Sec. (2.2). For better demonstrating such a parabola characteristic, the density axial profiles of cations and anions simulated are still compared to a parabola function constructed and based on the two critical data points that are sampled from the simulated density curve of summed cations, in Fig. 3. As seen, the similarity between the simulated and predicted parabolas is satisfactory. The simulated electrons density of Fig. 1(b) decreases in the electropositive halo, and is constant in the electronegative core, again in accord to the theoretical prediction of Sec. (2.2) in Eq. (25). The simulated and constant electron density in the electronegative core validates the Boltzmann's balance approximation used for the electrons and anions therein in the theory of Sec. (2.2). In addition, the cations and anions densities are not monotonic in their radial profiles but the density bumps appear in the core, as shown in Figs. 1(a) and 2(b). In Fig. 4(a, b) the simulated net source of anions is observed to be negative in the location where the anions density is peaked, and in Fig. 4(c) the simulated plasma potential is flat in the core, implying the free diffusion of anions therein. Since the required conditions, negative chemical source and free diffusion, are satisfied, the anions are self-coagulated to form the density bump in the radial profile of them, according to the theoretical analysis of Sec. (2.3). The density bump in the radial profile of cations is given by the ambi-polar coagulation of electronegative plasma at the high electronegativity condition, like the ambi-polar diffusion of electropositive plasma that keeps the electrical neutrality. The simulated flat potential of core in Fig. 4(b) is given by the flat electrons density simulated therein in Fig. 1(b), as determined by the Boltzmann's relation of electrons.

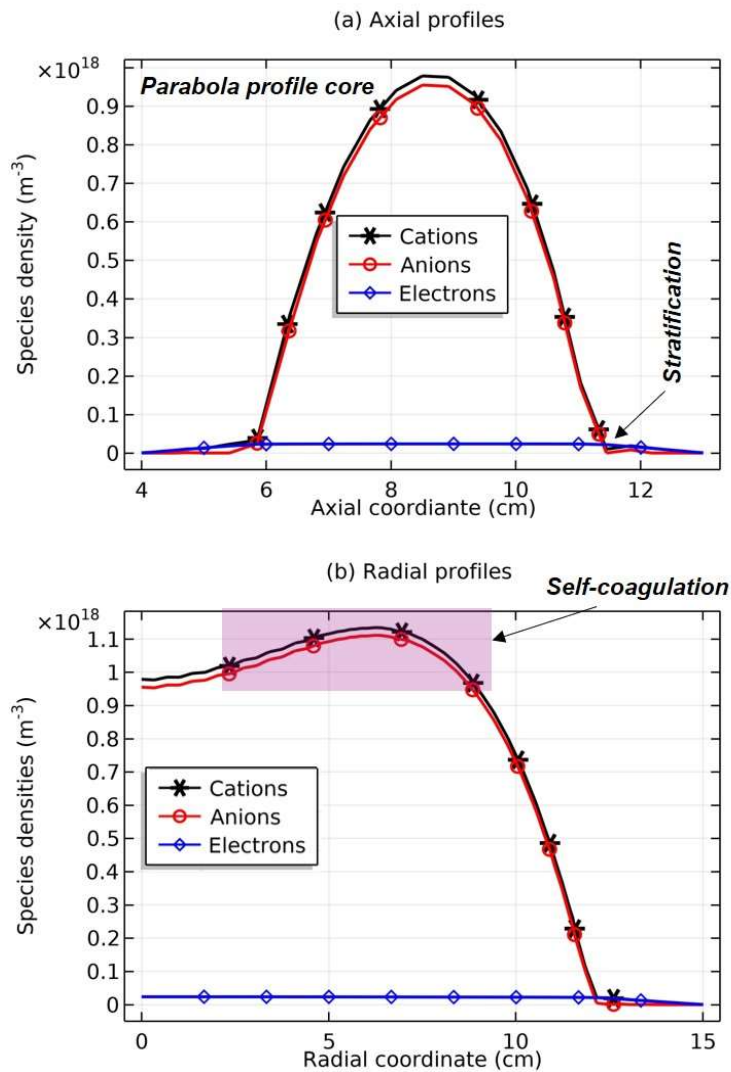

Figure 2. Simulated axial (a) and radial (b) profiles of the cations, anions, and electrons densities by the fluid model of Sec. (2.1). The discharge conditions of Ar/$SF_6$ ICP are 300W, 10mTorr and 10% $SF_6$ content and the simulated time is $0.1\,\text{s}$. In Panel (a) the density axial profiles of three plasma species are sampled along the central discharge axis, and in Panel (b) the density radial profiles of three plasma species are sampled at an axial position of $z = 8.5\,\text{cm}$.

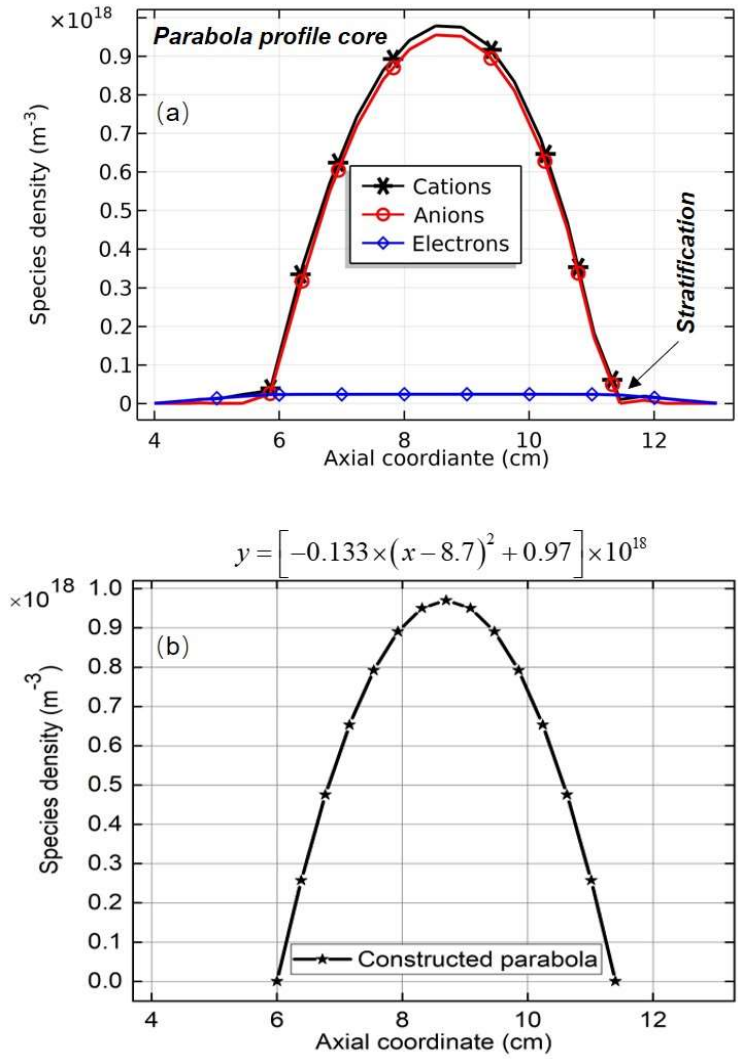

Figure 3. (a) Simulated axial profiles of species density and (b) constructed parabola function based on two critical points sampled from the simulated cations density curve, i.e., the peaked point with its coordinates, $(8.7, 0.97 \times 10^{18})$, and the truncated close-zero point with its coordinates, $(11.4, 0)$.

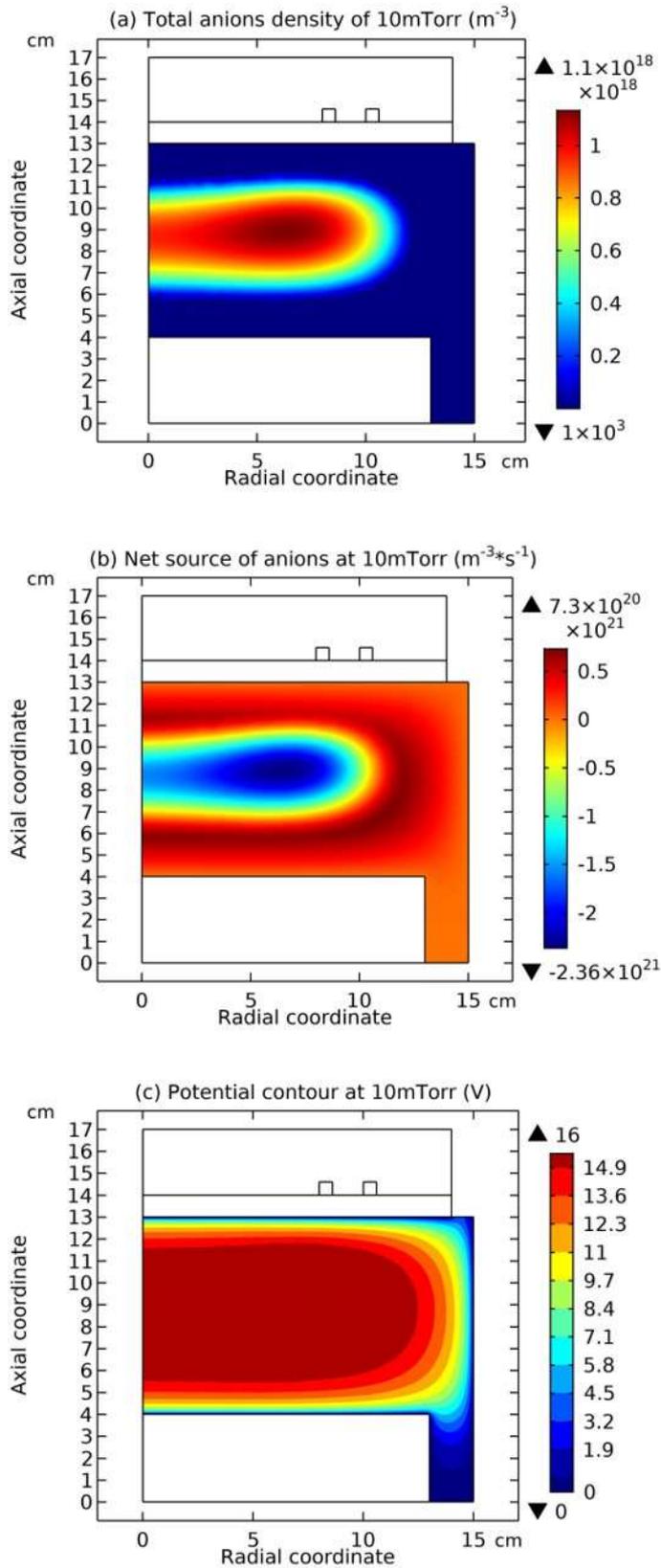

Figure 4. Simulated two-dimensional profiles of total anions density (a), net source of anions (b), and the plasma potential (c) by the fluid model of Sec. (2.1) in the Ar/SF$_6$ ICP, at 10 mTorr. The other discharge conditions and the simulated time are the same as in the Figs. 1 and 2.

## (3.2) The Boltzmann balance of anions and its possible origin, self-coagulation

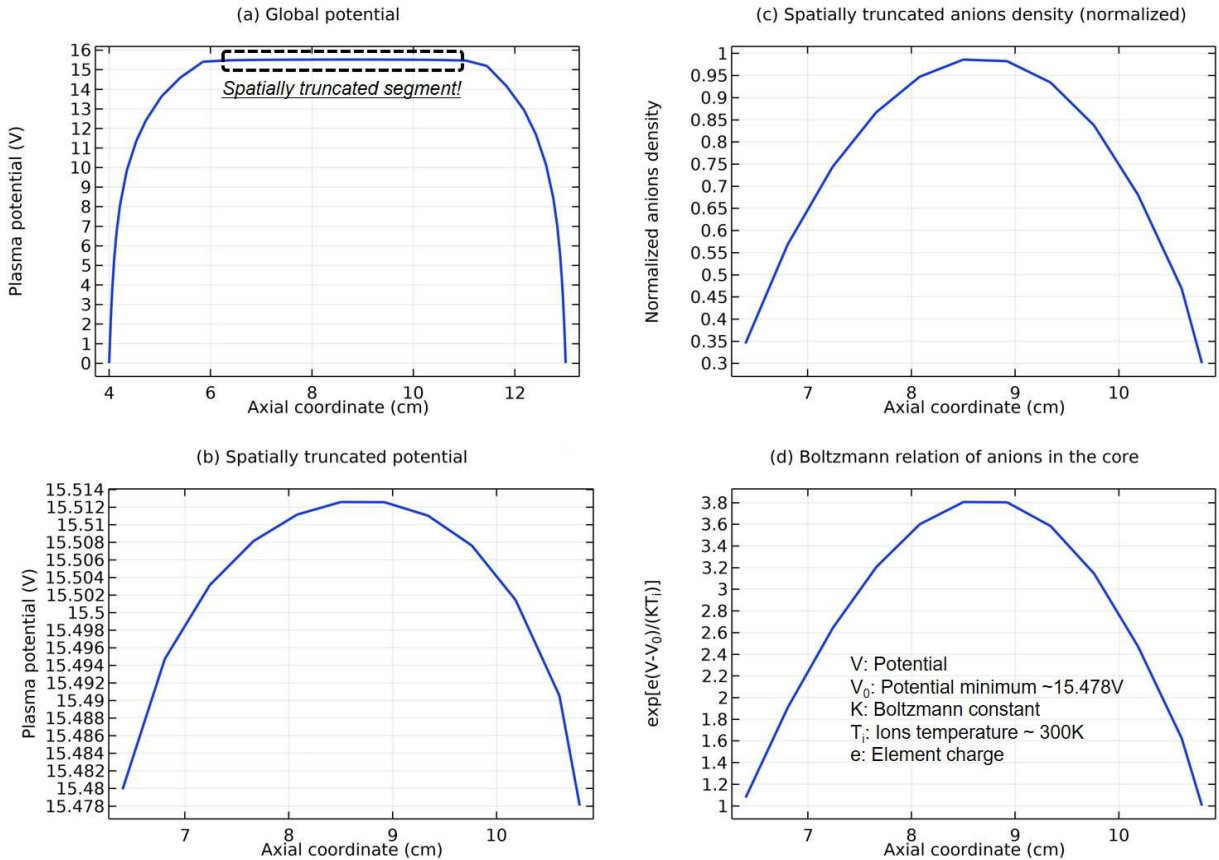

Figure 5. Simulated global and axial electrical potential of plasma (a), zoom in exhibition of the path of spatially truncated plasma potential (b), spatially truncated and normalized anions density of the same path (c) and the Boltzmann's relation of anions in the path (d). The discharge conditions of the Ar/SF$_6$ ICP and the simulated time are the same as in the Figs. 1 and 2.

As mentioned before, since the electrons density is in the Boltzmann's balance, the electrical potential of the plasma that is plotted globally and axially in Fig. 5(a) has the similar distribution as the electrons density shown in Fig. 1(b), that is, constant in the core but continually decreasing in the halo in a large potential unit, Volt, as observed from the vertical axis. As zooming in the potential and seen from the Figs. 5(a,b), in a truncated path as selected, the originally constant electrical potential in the core now has its definite profile, as demonstrated in a small range of potential, i.e., from $15.478 \text{ V}$ to $15.514 \text{V}$. In Fig. 5(c), the normalized anions density in this path is plotted, i.e., $\frac{n_-}{n_{-0}}$, and in Fig. 5(d) a certain function, $\exp\left[e(V-V_0)/KT_i\right]$, is constructed based on the tiny change of the potential as shown in Fig. 5(b) and on the room temperature of anions as assumed, that is, $T_i = 300 \text{ K}$. Here, $V_0$ is the potential minimum, $K$ is the Boltzmann's constant, and $e$ is the unit element charge, as illustrated in Fig. 5(d). As compared, the similarity between the two profiles of Fig. 5(c, d) indicates the anions satisfy the Boltzmann's balance, i.e.,

$$\frac{n_-}{n_{-0}} \sim \exp\left[e(V-V_0)/KT_i\right].$$

From the electromagnetics, it is known the potential and energy are exchangeable through the element charge. Therefore, the tiny potential change of the path shown in Fig. 5(b) can be described by the thermodynamic unit, Kelvin. The concrete processes are $\Delta = 15.514 - 15.478 = 0.036$ V and $0.036 \text{ eV} \sim \frac{0.036 \times 1.6 \times 10^{-19}}{1.38 \times 10^{-23}} \sim 417 \text{ K}$. In the formulae, the symbol, eV, is called the electronvolt. It is an energy unit, responsible for turning the potential into the energy. The electron charge, $1e = 1.6 \times 10^{-19}$ C, and the Boltzmann constant, $K: 1.38 \times 10^{-23}$ J/K, are utilized in the above formula to gives rise to the total variation of potential energy in that selected path of the core, about $417$ K. It is seen that the potential in the Boltzmann's balance of anions is in the room-temperature range, which is hence a weak potential barrel.

It is noted that the Boltzmann's balance of anions given by the fluid model simulation here was already validated by the experiments in Ref. [16] and widely used in the early analytical works of Refs. [11, 17-20, 22]. However, the origin of it is not clear yet. Herein, by means of the dimensional analysis of drift flux divergence and self-coagulation formulae, and by means of the characteristic of derivative of delta type density profile representing the coagulated structure, it is believed that the self-coagulation plays the role of drift that establishes the Boltzmann's balance of species, at the help of free diffusion.

$$\vec{\Gamma}_d = -\mu_- n_- \vec{E}_{eff}. \tag{44}$$

$$\nabla \cdot \vec{\Gamma}_d = -\mu_- \nabla n_- \cdot \vec{E}_{eff} - \mu_- n_- \nabla \cdot \vec{E}_{eff} = -\mu_- \nabla n_- \cdot \vec{E}_{eff} - \mu_- n_- \frac{\rho_{eff}}{\varepsilon_0}. \tag{45}$$

$$\nabla n_- \cdot \vec{E}_{eff} = \nabla \delta(\vec{r}) \cdot \vec{E}_{eff} = \begin{cases} 0, & \text{for } r \neq 0 \\ non-existing, & \text{for } r = 0 \end{cases}. \tag{46}$$

$$[\mu_-] = \frac{m^2}{V \cdot s}, \quad [n_-] = \frac{1}{m^3}, \quad [\rho_{eff}] = \frac{C}{m^3}, \quad [\varepsilon_0] = \frac{C}{m \cdot V},$$

$$[\mu_- n_- \frac{\rho_{eff}}{\varepsilon_0}] = \frac{1}{m^3 \cdot s} = [-n_- v_{rec}]. \tag{47}$$

$$-D_- \nabla^2 n_- - (-n_- v_{rec}) = 0,$$
$$-D_- \nabla^2 n_- - \mu_- n_- \nabla \cdot \vec{E}_{eff} = 0. \tag{48}$$

In Eq. (44), the drift flux of anions is assumed based on an *effective* electric field, and in Eq. (45) the divergence of such a flux is calculated, which is divided into two parts. The first part correlates the derivative of anions density, which will either be zero or not exist, at the assumption of delta profile, as shown in Eq. (46). For the second part of flux divergence, the Poisson's equation is used to transfer it to be a product of anions density, $n_-$, and an *effective* charge density, $\rho_{eff}$, together with the two constants, $\mu_-, \varepsilon_0$, as shown at the end of Eq. (45). After this transform, the expression of the second part of drift flux divergence is like that of the negative source in Eq. (31) of Sec. (2.3). The dimensional analysis of Eq. (47) further verifies the similarity between the second part of drift flux divergence and the negative source. So, the Quasi-Helmholtz equation in Eq. (31) is rewritten here in Eq. (48), and the negative source of it correlates directly to an *effective* drift flux. This is reasonable since the recombination of ions can be treated as the Coulomb's potential energy of cations and anions that attracts each other, as illustrated in Ref. [1]. It is stressed that only when the medium of flux is self-coagulated and the first part of drift flux divergence attenuates, this correlation exists, which implies that the self-coagulation provides the precursor of driving force for the electronegative plasma to evolve out the Boltzmann's balance of anions, i.e., the real and weak potential barrel shown in Fig. 5(b).

## (3.3) Investigation on the double layer
### (a) Its transferred models

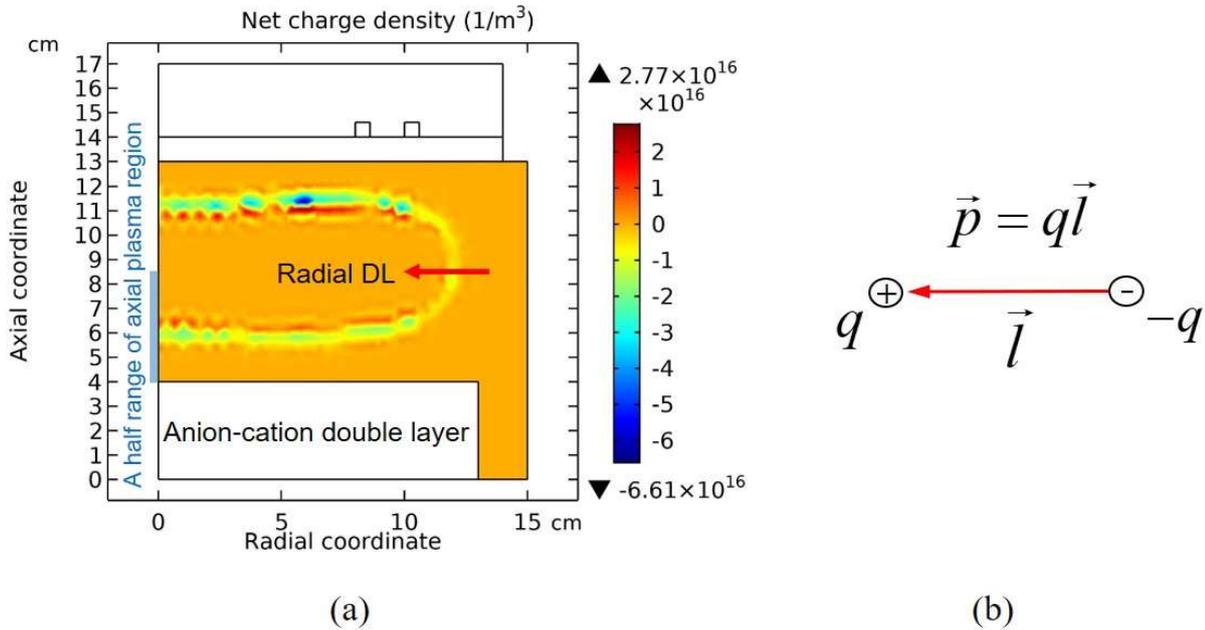

(a)                           (b)

Figure 6. (a) Simulated two-dimensional profile of net charge density, by the fluid model of Sec. (2.1) in the Ar/SF$_6$ ICP, and (b) transferred dipole model of double layer, abbreviated as DL, in the radial direction. The discharge conditions and the simulated time are the same as in the Figs. 1 and 2. The blue bar and corresponding text, a half region of axial plasma region, in Panel (a) will be used in Fig. 10(a).

    Besides for the stratification, the core parabola profile, and the coagulation, the self-consistent fluid simulation still shows that a double layer appears at the interface of stratification, which agrees with the simple fluid model prediction that includes the Poisson's equation in Refs. [17-20, 22]. In Fig. 6(a), the simulated two-dimensional profile of net charge density in the Ar/SF$_6$ ICP is shown. Upon comparing Fig. 6(a) to Fig. 1, it is seen that at the interface of the electronegative core plasma

and the electropositive halo plasma, the double layer structure, i.e., one layer of negative charge adjacent to one layer of positive charge, is appeared. In Fig. 6(b), this double layer is modelled as a dipole anti- along the direction of plasma transport, i.e., from the halo to core. Besides, it is seen from the figure that the two layers of charge are very close to each other. So, in Fig. 7, the spatial distribution of the electric field intensity of such a dipole at the limit of dipole distance tending to zero is given. It is seen that the field is strong at the dipole center but tends to zero at other points. The concrete mathematic process is presented in Eqs. (49-52). In both the Fig. 7 and the related equation set, the point, $O$, is the dipole center and the point, $A$, is an arbitrary *far-field* point of the space. $E_O, E_A$ are the dipole fields of points, $O, A$, respectively. Herein, $l$ is the distance of dipole moment and $q$ is the dipole charge. Hence, $\vec{p} = q\vec{l}$ and it is the dipole moment. $r$ is the distance of dipole center to the point, $A$, and $\vec{r}_0$ is the corresponding unit vector. $\varepsilon_0$ is the vacuum permittivity. The field intensities at *far-field* points tend to zero because of the counteracting effect of positive and negative charges while the field intensity at the dipole center is infinitely large because of the localization effect, at the limit of $l \to 0$.

$$E_O = \frac{2q}{\pi \varepsilon_0 l^2}, \tag{49}$$

$$\lim_{l \to 0} E_O = \lim_{l \to 0} \frac{2q}{\pi \varepsilon_0 l^2} = \infty, \tag{50}$$

$$\vec{E}_A = \frac{1}{4\pi \varepsilon_0 r^3} \left[ -\vec{p} + 3(\vec{r}_0 \cdot \vec{p}) \vec{r}_0 \right], \tag{51}$$

$$l \to 0, \vec{p} = q\vec{l} \to \vec{0}, \Rightarrow \vec{E}_A \sim \vec{0}. \tag{52}$$

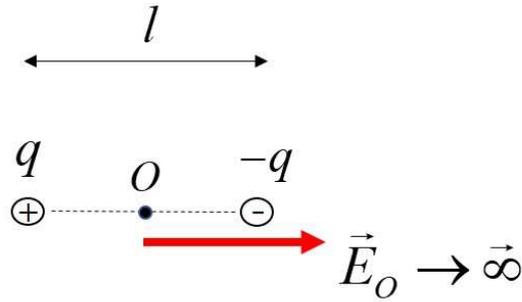

(a) Electric field intensity of dipole moment at the dipole center

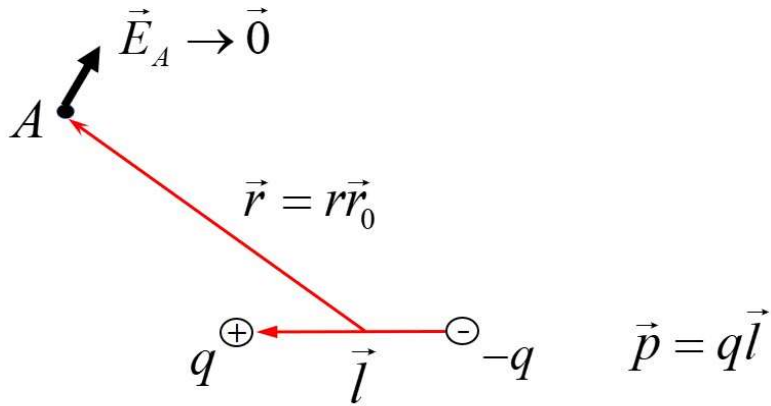

(b) Electric field intensity of dipole moment at arbitrary *far-field* point of the space

Figure 7. Spatial distribution of the electric field intensity of the transferred dipole model at the limit of $l \to 0$ at (a) the dipole center and (b) arbitrary *far-field* point. Herein, $l$ is the distance of dipole moment.

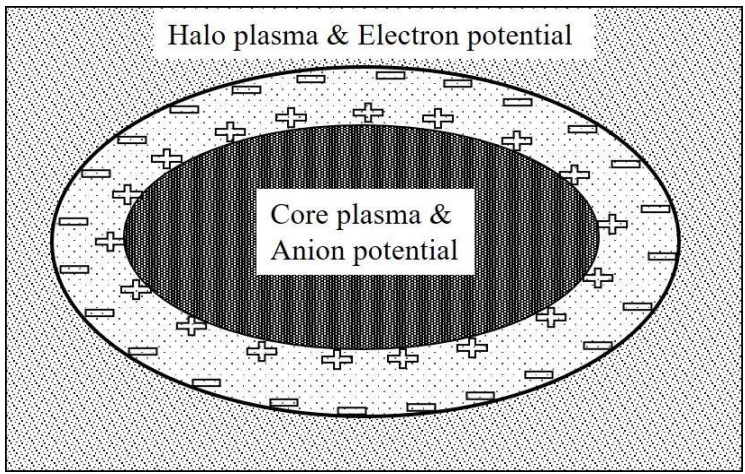

Figure 8. Transferred ellipsoid capacitor model of double layer

    Considering the three-dimensional geometry of cylindrical chamber and the axial symmetry, the double layer, when seen along the whole interface of the electronegative core plasma and the electropositive halo plasma, can be transferred to a capacitor of the ellipsoid, as shown in Fig. 8. Herein, the electron potential in the external halo plasma is meant the ambi-polar diffusion potential in a unit of electronvolt, and the anion potential in the internal core plasma is meant the weak

potential in a unit of Kelvin that supports the Boltzmann's balance of anions in Fig. 5 of Sec. (3.2). Different with the CCP source that is radio frequency vibrated, the transport of ICP source is more like a direct current (DC) plasma source in the radial and axial directions. The capacitor is known to be able to block the direct current, which means that the electropositive halo is separated from the electronegative core, as seen from Fig. 9 where a transferred circuit model of such an ICP source is given. The stratification of the Ar/SF$_6$ ICP discharge structure through the capacitor model of the double layer is triggered by the external power source, as illustrated in this figure. As seen next, the electrical field and potential properties of the transferred dipole and capacitor models of double layer are both validated by the simulation.

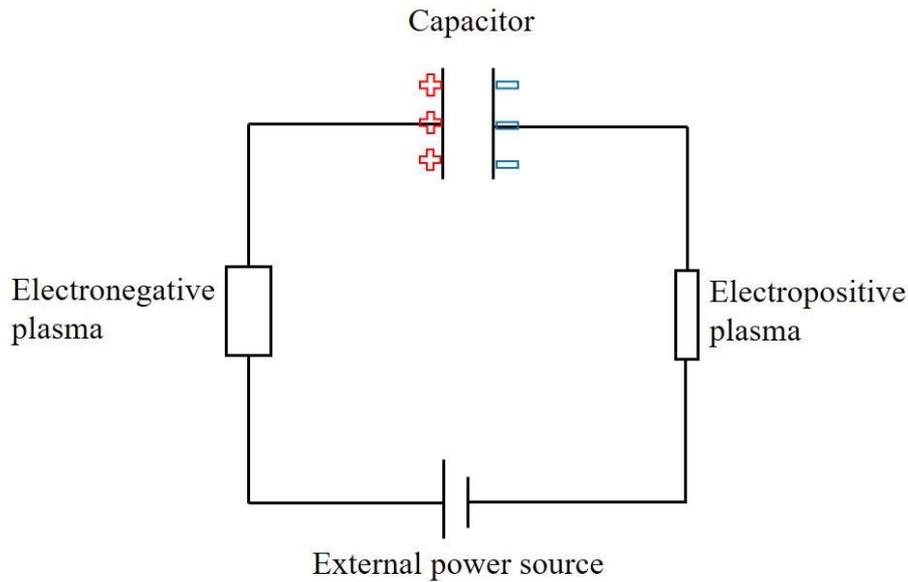

Figure 9. Transferred circuit model of the Ar/SF$_6$ ICP.

**(b) Double-valued properties of plasma edge potential and cations flux, at the DL**

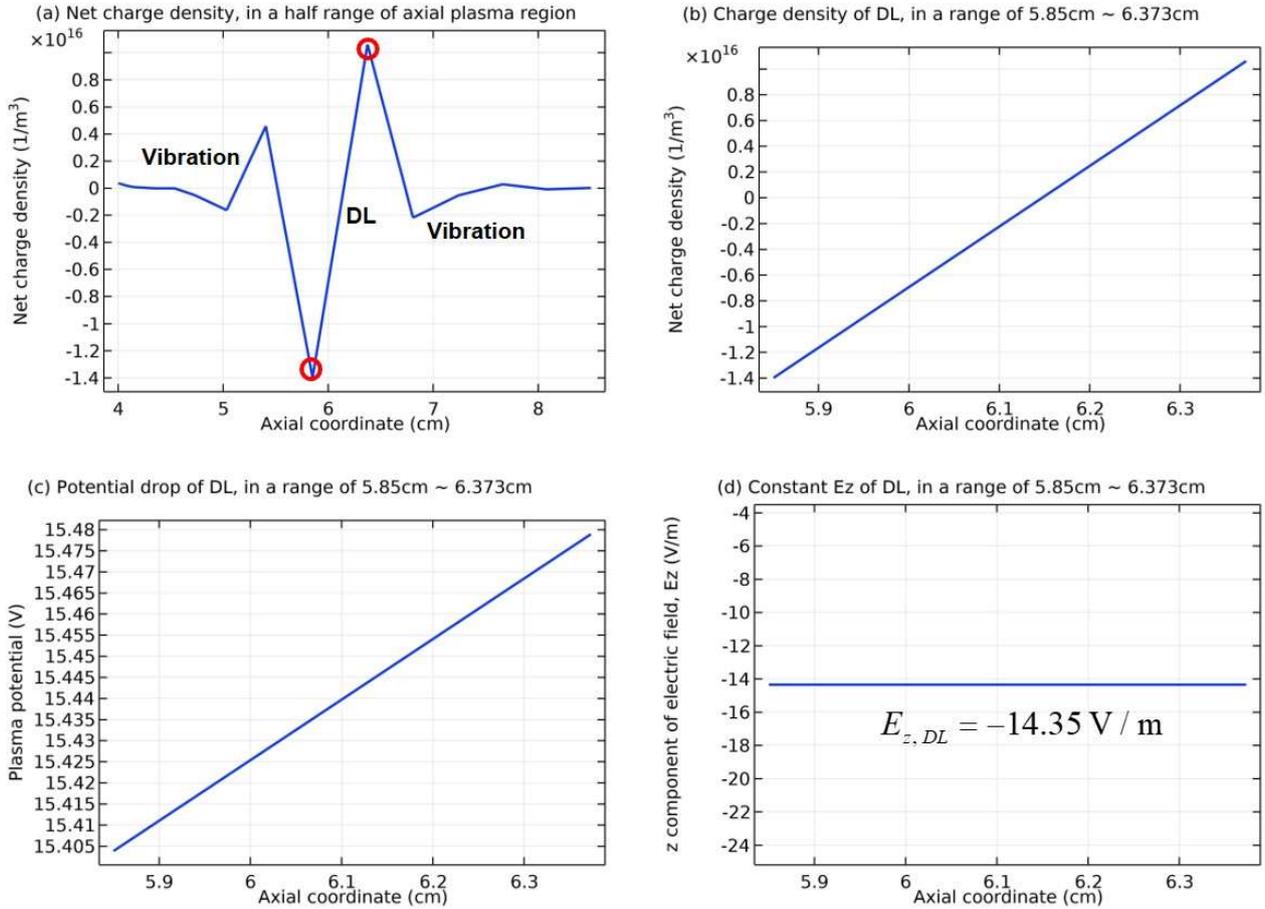

Figure 10. (a) Net charge density in a half range of axial plasma region (see Figure 6(a) for reference), (b) charge density of double layer in an axial range from 5.85 to 6.373 cm, (c) potential drop of double layer still in this range, and (d) display for the constant axial electrical field component of double layer, $-14.35 \text{ V/m}$, in the same range, all along the central axis of discharge and given by the fluid simulation of Sec. (2.1). The discharge conditions of the Ar/SF$_6$ ICP and the simulated time are the same as in the Figs. 1 and 2. In Panel (a), the region of DL is marked and at the two sides of it, the ionic and acoustic vibrations appear, which will be analyzed in Sec. (3.3c)

In this section, the terms, plasma edge potential and flux, were illustrated in the literature review of Sec. I. They were defined in the early analytical works of Sheridan *et al* in Refs. [19, 20, 22]. In these works, both the method based on the plasma approximation, i.e., without the Poisson equation, and the one that includes the non-neutral region, i.e., with the Poisson equation, were used. The applications of the two methods and the correlation of them revealed that at the plasma edge, i.e., the interface between the electrically neutral plasma and the non-neutral region, both the potential and the flux of cations are double-valued, and when the flux magnitude of cations with the high edge potential was larger than the one with the low edge potential, the double layer appeared at the plasma edge. Our fluid simulation perfectly describes this picture in Figs. 10-12, and moreover, it finds out the behind physics for explaining such a trend related to the double layer in the electronegative plasma sources.

In Fig. 10(a), the simulated net charge density by the fluid model of Sec. (2.1), in a half range of axial plasma region along the central axis of discharge, is given, where the double layer and the related vibrations are presented. In Fig. 10(b,c,d), the net charge density, the potential drop, and the axial electric field component in an axial range of double layer location, i.e., from 5.85 to 6.373 cm, are presented. As seen before, the double layer in Fig. 10(b) consists of two charge layers with different polarities. The potential in Fig. 10(c) drops with the direction oriented to the halo and it, called as the plasma edge potential, is indeed double-valued in this non-neutral region, in accord to the theoretical predictions of Refs. [19, 20, 22]. The axial electrical field of double layer in Fig. 10(d) is constant and its magnitude is $-14.35$ V/m, which agree qualitatively with the dipole and capacitor model predictions on the electric field properties in Sec. (3.3a). In Fig. 11, the general trends of the axial electrical field and the potential for the regions of the sheath, the electropositive halo, the double layer, and the electronegative core, are presented. From this figure, the intensities of electrical field in the different regions can be ordered, i.e., sheath, electropositive halo, double layer, and electronegative core. Similarly, the potential is ordered as well, but reversely as noticed, i.e., electronegative core, double layer, electropositive halo, and sheath. In a word, the potential consecutively decreases from the inner core to the outer sheath.

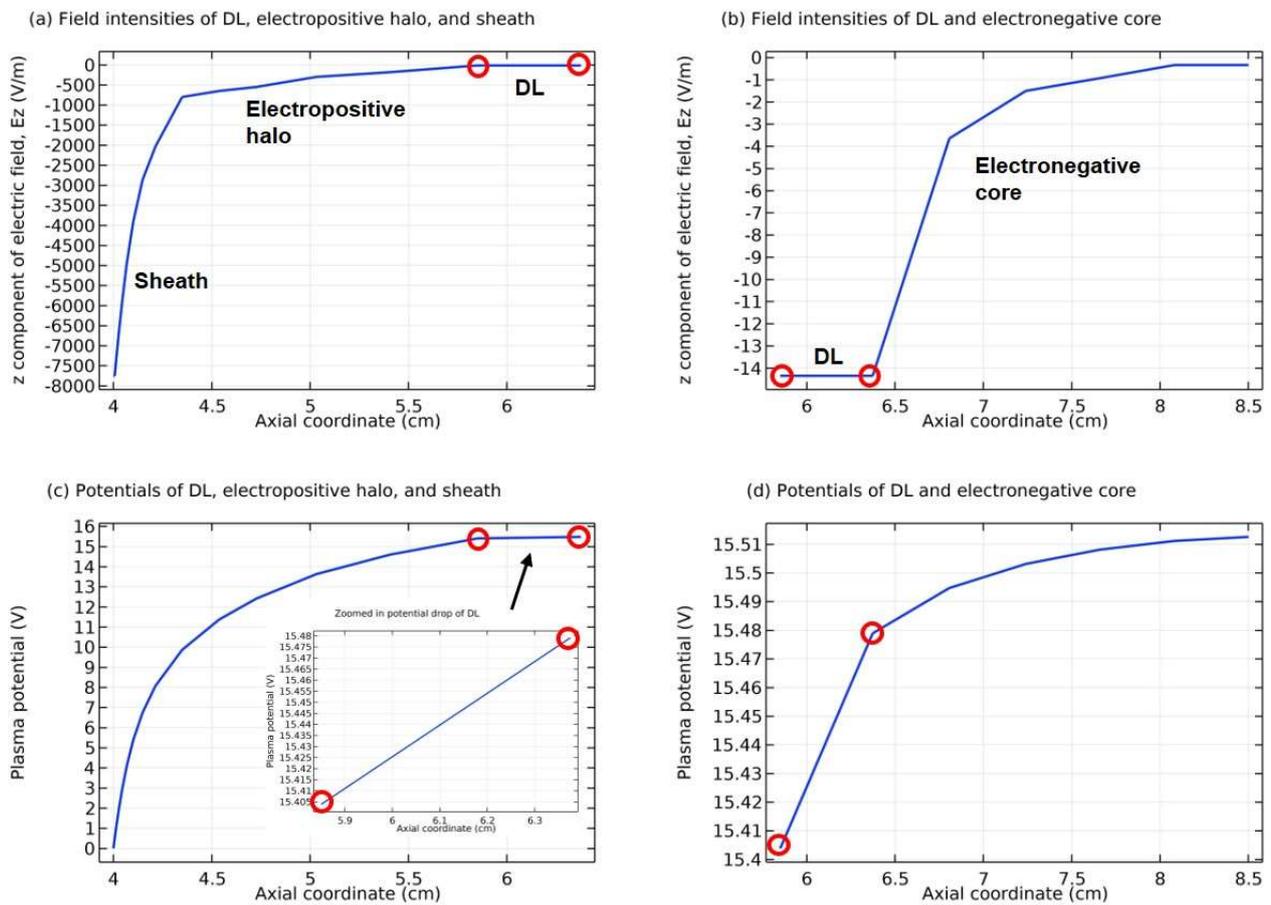

Figure 11. Axial profiles of axial electrical field component for (a) the regions of double layer, electropositive halo, and sheath and for (b) the regions of double layer and electronegative core, and still the axial profiles of potential for (c) the regions of double layer, electropositive halo, and sheath and for (d) the regions of double layer and electronegative core, all along the central axis of discharge and given by the fluid simulation of Sec. (2.1). The discharge conditions of the Ar/SF$_6$ ICP and the simulated time are the same as in the Figs. 1 and 2. In Panels (a) and (b) the electrical field intensity of double layer is truly constant as demonstrated in Fig. 10(d), and in Panel (c) the potential of double layer is not flat as illustrated in the inserted sub-figure of Panel (c) and in Panel (d) when the small ranges of potential value are used for the vertical axes.

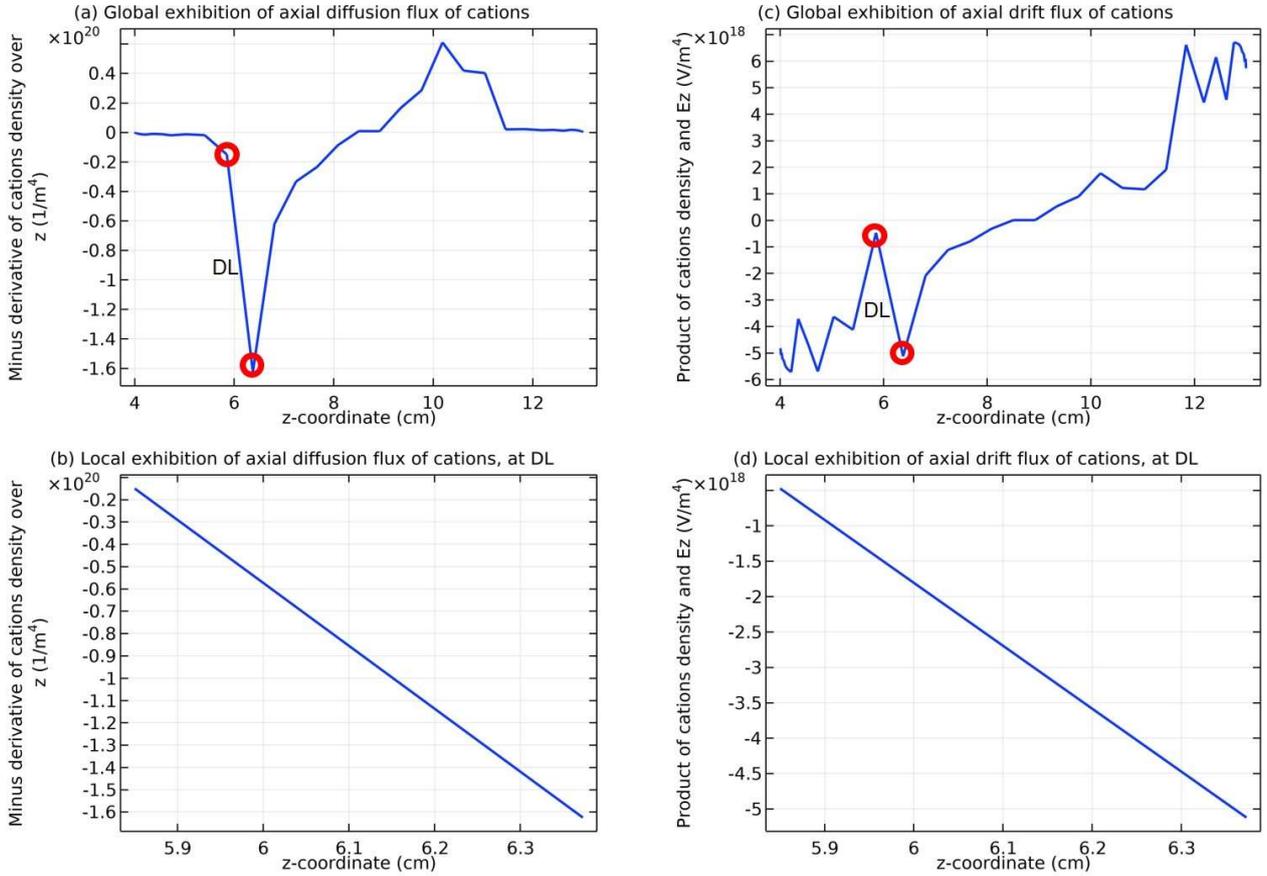

Figure 12. (a) Global exhibition of axial *diffusion* flux of cations and (b) local exhibition of it at the double layer, and (c) global exhibition of axial *drift* flux of cations and (d) local exhibition of it at the double layer, all along the central axis of discharge and given by the fluid simulation of Sec. (2.1). The discharge conditions of the Ar/$SF_6$ ICP and the simulated time are the same as in the Figs. 1 and 2. The definitions for the diffusion and drift fluxes herein are as follows, i.e., the axial *drift* flux of cations is represented by the product of cations density and axial electrical field and the axial *diffusion* flux is represented by the minus derivative of cations density over the axial coordinate. See the dimensions of the two representative fluxes for reference. The qualitative trends of two fluxes, not the quantitative values of them, are emphasized in this figure.

    The axial profiles of representative axial diffusion and drift fluxes of cations are plotted both globally in the whole range in Fig. 12(a, c) and locally in the small range of double layer in Fig. 12(b, d). In Fig. 12(a, c) the positions of double layers in the flux curves are marked and both the diffusion and drift fluxes of cations are double-valued in the range of double layer, again in accord to the theoretical predictions of Refs. [19, 20, 22]. More clearly in Fig. 12(b, d) the magnitudes of the two types of cations fluxes near the electronegative core, with the high edge potential as seen from Figs. 10(c) and 11(d), are indeed larger than the ones near the electropositive halo, with the low edge potential as seen from Figs. 10(c) and 11(c), again in agreement with the theoretical predictions of Refs. [19, 20, 22] on the relation of plasma edge potential and flux that are both double-valued. It is originated from the simulated facts that the potential consecutively drops from the core to the halo, through the double layer, as shown in Fig. 11(c,d) and the flux magnitudes in

the core are naturally larger than the ones in the halo where the density of cations summed is several orders lower, as shown in Fig. 2(a).

### (c) Analysis on the double layer origin and consequence

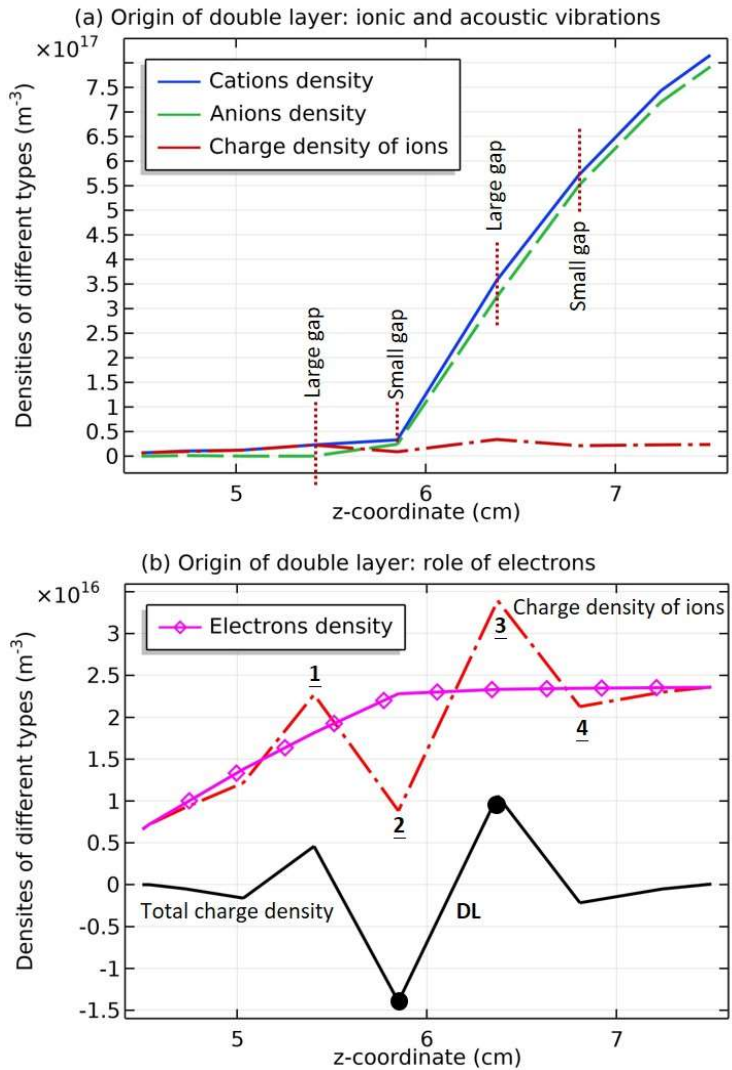

Figure 13. (a) Axial profiles of cations density (blue line), anions density (green line) and charge density of ions (red line), i.e., the deduct of cations and anions densities, and (b) axial profiles of charge density of ions (red line), electrons density (purple line) and the total charge density (black line), i.e., the deduct of cations density and the sum of anions and electrons densities, all along the central axis of discharge and given by the fluid simulation of Sec. (2.1). The discharge conditions of the $Ar/SF_6$ ICP and the simulated time are the same as in the Figs. 1 and 2.

In Fig. 13, the axial profiles of cations density, anions density, charge density of ions, electrons density and the total charge density simulated are shown. It is seen in Fig. 13(a) that the double layer hidden in the curve of charge density of ions is originated from the ionic and acoustic vibrations of cations and anions densities, and in Fig. 13(b) that the electrons are less related to the formation of double layer and the role of them is to just shift down the curve of ionic charge density. The double layer marked in Fig. 13(b) is formed at the interface between the core and halo, where the ionic and acoustic vibration is the strongest. Away from the interface, these vibrations are smoothed by the diffusion of ionic species, as illustrated in Ref. [25].

The double layer simulated is formed by the anions sheath in the side of electropositive halo (which gives rise to the anion net charge density) and the cations sheath in the side of electronegative core (which gives rise to the cations net charge density). The anions sheath is formed by the pushing of ambi-polar potential in the halo. In our opinion, the anions sheath is formed with the thermal velocity of anions that plays the role of Bohm's sheath criterion, since the pushing of strong ambi-polar potential of halo on the anions are against role, instead of collective interaction. This is proven by our convergence that will be shown in future, in which the anions net charge density is given at the discharge initial. The cations sheath in the core side is formed by the accelerating of weak ambi-polar potential of core onto the cations. The Bohm's velocity of cations at the strong electronegativity is expressed as $u_B = \left[\dfrac{eT_e(1+\alpha)}{M_+(1+\gamma\alpha)}\right]^{1/2}$ [11, 27], wherein $\alpha \sim 100.0$ is the electronegativity and $\gamma \sim 100.0$ is the ratio between the electron and anion temperatures, at the quasi cold ion approximation. As seen, at the above conditions, the reduced Bohm's velocity of cations is its thermal velocity, $u_B \sim \left[\dfrac{eT_i}{M_+}\right]^{1/2}$ at the condition of non-thermal equilibrium, $\dfrac{T_e}{T_i} = \gamma \sim 100.0$, at which the anions are cold enough, as the early analytic works predicted. As seen, this reduced Bohm's velocity can be achieved by the acceleration of weak potential barrel given in Sec. (3.2).

In the conventional steady state double layer picture, four groups of species, the trapped electrons and ions and the free electrons and ions, are needed [36]. Here in the electronegative plasma, the trapped anions and cations and free anions and ions are needed. The trapped anions mean the anions that are generated through the attachments in the core and confined therein by the double layer electric field that is infinitely large as shown in Fig. 7, and the trapped cations mean the cations that are generated through ionizations in the halo and confined therein by the double layer electric field. The free anions mean the anions that are generated through the attachments in the halo and can be accelerated through the double layer field, hence forming the upstream anions beam. The free cations mean the cations that are generated through the ionizations in the core and can be accelerated through the double layer field, hence forming the downstream cations beam. The coexistence of anions and cations beams can excite the longitudinal acoustic wave. However, this acoustic wave cannot transmit in the plasma media since the wave will be strongly suppressed by the Landau damping of ions, when the wave phase velocity approaches to the thermal velocity [2]. In the electronegative plasma, the acoustic wave velocity is obtained as $\dfrac{\omega}{k} = \left[\dfrac{eT_- + \gamma_i eT_i}{M_+}\right]^{1/2} \equiv v_s$ at the anions Boltzmann's relation, not the electron's, as illustrated in Sec. (2.4). Herein, $T_- = T_i = 300\text{ K}$, and $\gamma_i \sim 3$ is the adiabatic coefficient of cations and its value can be selected as 3, at the condition of collisionless case. It is noticed that the collisionless condition here is validated because the double layer and acoustic wave always appear in the collisionless plasma, as reported in Refs. [36, 37]. The wave phase velocity obtained equals to $2\left[\dfrac{eT_i}{M_+}\right]^{1/2}$, in the order of ionic thermal velocity. The angular frequency of acoustic wave is given as the ionic Langmuir's vibration, $\omega_i = \left[\dfrac{n_i e^2}{M_+ \varepsilon_0}\right]^{1/2}$, herein $n_i \sim 2.5 \times 10^{16}$ m$^{-3}$ is the certain cations density at the double layer position given by the fluid simulation, as illustrated in Fig. 13(a). Based on the wave velocity and ionic frequency, the calculated wave length in Eq. (53) is within the order of centimeter, in good accord to Fig. 13(b), in which more or less two wave lengths are experienced from 5.0 cm to 7.0 cm.

$$\lambda_s = \frac{2\pi v_s}{\omega_i} = \frac{4\pi (eT_i/M_i)^{1/2}}{\left(\dfrac{n_i e^2}{M_i \varepsilon_0}\right)^{1/2}} = 4\pi \left(\frac{T_i \varepsilon_0}{en_i}\right)^{1/2} = 1.02 \text{ cm.} \quad (53)$$

It is noted that the double layer is formed by the pair of cations and anions, and the electrons do not attend this process since they are light and thermal and it is easy for them to cross the potential barrel of double layer shown in Fig. 10(c), ~ 0.075V. Actually, the suggested mechanism of damped acoustic wave herein given by the double layer at the equivalent cations and anions temperatures is applied to the case of Ref. [25] as well, where the oscillation of double layer is suppressed after considering the diffusion of cations, which reduces the discrepancy of anions and cations temperatures, i.e., approaching to the quasi- cold ion approximation. The temperature of cations is otherwise zero while the temperature of anions is the room temperature, which suits for the transmission of acoustic wave, without the Landau's damping.

## VI. Conclusion and further remarks

In this work, the early analytic works of electronegative discharge structure are systematically analyzed. Their advantages and disadvantages are elaborated and the significance of self-consistent fluid simulation at the quasi- cold ion approximation on revealing the discharge hierarchy at low pressure is clarified.

The outline of the hierarchy is as follows. The ambi-polar diffusion processes of both the electropositive plasma and electronegative plasma are faster at the low pressure, i.e., with the small timescale as compared with the attachment [15], and hence the discharge is stratified into the core and halo. In the electronegative core, the ambi-polar self-coagulation of ions occurs, which leads to the anions Boltzmann's balance and the parabola profiles of ions. This finding validates the guess of Ref. [15], and meanwhile offers more physics for it. At the interface of core and halo, the double layer occurs due to the formation of cation and anion sheaths, at the necessary conditions of quasi- cold ion approximation and high electronegativity. The longitudinal acoustic wave is first triggered by the cation and anion beams created by the double layer potential drop, and then suppressed by the Landau's damping because the wave phase velocity approaches to the ionic thermal velocity, again at the quasi- cold ion approximation. This explains the math fact of Ref. [25] that the oscillation of double layer disappears after including the free diffusion of cations. In addition, the simulation also explains the relation of potential and flux at the double layer that was found in the early analytic works.

The present fluid simulation that describes the discharge hierarchy needs to be compared with the more self-consistent fluid simulation that considers the Ohm's heating scheme of ions [30]. Certain characteristics in the hierarchy, such as the self-coagulation, the uniform profiles of electrons density and plasma potential in the core, the double layer electric field intensity, the coexistence of thermal and directed ions, and the related acoustic wave that's damped, probably find their potential applications in future, such as the free fusion, the uniform electron source, the precursor for high energy ionic beams, and the safe information communication through damped wave.